\documentstyle[amssymb,epsfig]{fbssuppl}
\begin{document}
\title{Integral equations for three-body Coulombic
resonances\footnote{This work is
dedicated to the 60th birthday of
Prof. W. Gl\"ockle.}}
\author{Z.\ Papp${}^{1}$, I.\ N.\ Filikhin${}^{2}$
 and S.\ L.\ Yakovlev${}^{2}$}
\institute{
${}^{1}$Institute of Nuclear Research of the
Hungarian Academy of Sciences, \\
P.O. Box 51, H--4001 Debrecen, Hungary \\
${}^{2}$ Department of Mathematical and Computational Physics, \\
Sankt-Petersburg State University, \\
198904 Sankt-Petersburg, Petrodvoretz,
Ulyanovskaya Str. 1, Russia}
\date{\today}
\maketitle

\begin{abstract}
\noindent
We propose a novel method for calculating resonances in three-body
Coulombic systems. The method is based on the solution of the
set of Faddeev and Lippmann-Schwinger integral equations,
which are designed for solving the three-body Coulomb problem.
The resonances of the three-body system are defined
as the complex-energy solutions of the homogeneous
Faddeev integral equations.
We show how the kernels of the integral equations should be
continued analytically in order that we get resonances.
As a numerical illustration a toy model for the three-$\alpha$
system is solved.
\end{abstract}
%\vspace{0.5cm}

\section{Introduction}

For three-body systems the Faddeev equations are the fundamental equations.
After one iteration they possess connected kernels,  consequently they are
effectively Fredholm integral equations of second kind. Thus the Fredholm
alternative applies: at certain energy either the homogeneous or inhomogeneous
equations have solutions. Three-body bound states correspond to
the solutions of the homogeneous Faddeev equations at real energies,
resonances, as usual in quantum mechanics,
are related to their complex-energy solutions.

The situation becomes more complicated if the Coulomb potential
enters into the game. The Faddeev equations were derived for short range
interactions and if we simply plug-in a Coulomb-like potential they become
singular.
The solution has been formulated
 in the Faddeev-Merkuriev theory \cite{fm-book}
on a mathematically sound and elegant way via
integral equations with connected (compact) kernels and
configuration space differential equations 
with asymptotic boundary conditions.

Recently, along the concept of a ''three-potential'' picture,
 a novel method has been proposed
for treating the three-body Coulomb problem
via solving the set of Faddeev-Noble and Lippmann-Schwinger
integral equations in Coulomb--Sturmian-space representation.
The method was elaborated first
for bound-state problems \cite{pzwp} with repulsive Coulomb plus nuclear
potential, then it was extended for calculating $p-d$
scattering at energies below the breakup threshold \cite{pzsc}.
In these calculations an excellent agreements with the results of other well
established methods were found and
the efficiency and the accuracy of the method were demonstrated.
Also atomic  bound-state problems with attractive Coulomb
interactions were considered \cite{pzatom}.
The mathematical validity of the procedure, in the
context of the Faddeev-Merkuriev theory, has also been established \cite{pzsy}.

In this article we seek to extend this novel method for
calculating resonant states, the complex-energy solutions of the
homogeneous Faddeev-type integral equations. First, along the
method of Ref.\ \cite{pzwp}, we recapitulate the solution of the
homogeneous integral equations, then we show how to continue
analytically the Green's operators onto the unphysical sheet. As
an illustration of the feasibility  of this method
we calculate a model three-$\alpha$ system interacting via
short-range plus repulsive Coulomb interaction.

\section{Faddeev-Merkuriev integral equations}
The Hamiltonian of a three-body Coulombic  system reads
\begin{equation}
H=H^0 +W + v_\alpha+ v_\beta + v_\gamma,
\label{H}
\end{equation}
where $H^0$ is the three-body kinetic energy
operator, $W$ stands for the
three-body potential and $v_\alpha$ denotes the
Coulomb-like interaction in the subsystem $\alpha$.
We use throughout the usual
configuration-space Jacobi coordinates
$x_\alpha$  and $y_\alpha$.
Thus  $v_\alpha$ only depends on $x_\alpha$ ($v_\alpha=v_\alpha (x_\alpha)$),
while $W$ depends on both  $x_\alpha$ and $y_\alpha$ coordinates
($W=W(x_\alpha,y_\alpha)$).

The physical role of a Coulomb-like potential is twofold.
Its long-distance part modifies the asymptotic
motion, while its short-range part strongly correlates the
two-body subsystems.
Merkuriev proposed to split the potentials into short-range and long-range
parts in the three-body configuration
space via a cut-off function $\zeta$ \cite{fm-book},
\begin{equation}
v_\alpha^{(s)}(x_\alpha,y_\alpha)=
v_\alpha(x_\alpha) \zeta_\alpha(x_\alpha,y_\alpha),\end{equation}
and
\begin{equation}v_\alpha^{(l)}(x_\alpha,y_\alpha)=
v_\alpha(x_\alpha) [1- \zeta_\alpha(x_\alpha,y_\alpha) ].
\label{potm}
\end{equation}
The function $\zeta_\alpha$ is defined such that it separates the
asymptotic two-body sector $\Omega_\alpha$ from the rest
of the three-body configuration space.
On the region of $\Omega_\alpha$
 the splitting function $\zeta_\alpha$
asymptotically tends to $1$ and  on the complementary asymptotic region
of the configuration space it tends
to $0$. Rigorously, $\Omega_\alpha$ is defined as a part of the
three-body configuration
space where the condition
%%%%%%%%%%%%%%%%%%%%%
\begin{equation}
|x_\alpha| < a (1+|y_\alpha|/a)^\nu, \mbox{with} \ \ \ a>0,\  0<\nu < 1/2,
\label{oma}
\end{equation}
%%%%%%%%%%%%%%%%%%%%%
is satisfied. So, in $\Omega_\alpha$ the short-range part
$v_\alpha^{(s)}$ coincides with the
original  Coulomb-like potential $v_\alpha$
and in the complementary region vanishes, whereas
the opposite holds true for $v_\alpha^{(l)}$. From its construction
follows that if $a$ is chosen big enough
$\sum_{\alpha}v_\alpha^{(l)}$ in the three-body Hilbert space
does not support any bound state \cite{fm-book}.
This  phenomena is analogous to the observation that some special atomic
three-body systems does not have any bound states \cite{galindo}.
Note that for repulsive Coulomb interactions
one can also adopt Noble's approach \cite{noble}, which can be considered as
a special case of Merkuriev's splitting, where the splitting is
performed in the two-body configuration space.
Then $v_\alpha^{(l)}$ coincides with the whole Coulomb interaction and
$v_\alpha^{(s)}$ with the short range nuclear potential.

In the Faddeev procedure we spilt the wave function into
three components
\begin{equation}
|\Psi \rangle = |\psi_{\alpha} \rangle +
|\psi_{\beta} \rangle +|\psi_{\alpha} \rangle
\end{equation}
by applying the asymptotic filtering \cite{vanzani} 
operator 
\begin{equation}
|\psi_\alpha \rangle = G^{(l)} (z) v_\alpha^{(s)} |\Psi \rangle.
\end{equation}
Here $G^{(l)}$ is the resolvent of the long-ranged Hamiltonian
\begin{equation}
H^{(l)} = H^0 + W +  v_\alpha^{(l)}+
v_\beta^{(l)}+ v_\gamma^{(l)}
\label{hl}
\end{equation}
and $z$ is the complex energy parameter.
The wave-function components satisfy the homogeneous
Faddeev-Merkuriev integral equations
\begin{equation}
|\psi_{\alpha} \rangle= G_\alpha^{(l)} (z)
v^{(s)}_\alpha \sum_{\gamma\neq\alpha}
|\psi_{\gamma} \rangle,
\label{fn-eq}
\end{equation}
where $G^{(l)}_\alpha$ is the resolvent of the channel 
long-ranged Hamiltonian
\begin{equation}
H^{(l)}_\alpha = H^{(l)} + v_\alpha^{(s)}.
\label{hla}
\end{equation}
Merkuriev has proved that Egs.\ (\ref{fn-eq}) possess compact kernels 
for positive $E$ energies, and this property remains valid also for
complex energies $z=E-i\Gamma/2$, $\Gamma > 0$.

\section{Solution method}

We solve these integral equations
in Coulomb--Sturmian-space representation.
The Coulomb-Sturmian (CS) functions are defined by
\begin{equation}
\langle r|n \rangle =\left[ \frac {n!} {(n+2l+1)!} \right]^{1/2}
(2br)^{l+1} \exp(-b r) L_n^{2l+1}(2b r),  \label{basisr}
\end{equation}
with $n$ and $l$ being the radial and
orbital angular momentum quantum numbers, respectively, and $b$ is the size
parameter of the basis.
The CS functions $\{ |n \rangle \}$
form a biorthonormal
discrete basis in the radial two-body Hilbert space; the biorthogonal
partner defined  by $\langle r |\widetilde{n }\rangle=
r^{-1} \langle r |{n}\rangle$. 
Since the three-body Hilbert space is a direct product of two-body
Hilbert spaces, an appropriate basis
is the angular momentum coupled direct product (the possible other
quantum numbers are implicitly assumed)
\begin{equation}
| n \nu \rangle_\alpha =
 | n  \rangle_\alpha \otimes |
\nu \rangle_\alpha, \ \ \ \ (n,\nu=0,1,2,\ldots).
\label{cs3}
\end{equation}
With this basis the completeness relation
takes the form
\begin{equation}
{\bf 1} =\lim\limits_{N\to\infty} \sum_{n,\nu=0}^N |
 \widetilde{n \nu } \rangle_\alpha \;\mbox{}_\alpha\langle
{n \nu } | =
\lim\limits_{N\to\infty} {\bf 1}^{N}_\alpha.
\end{equation}
Note that in the three-body Hilbert space,
three equivalent bases belonging to fragmentation
$\alpha$, $\beta$ and $\gamma$ are possible.

In Ref.\ \cite{pzwp} a novel approximation scheme has been
proposed to the Faddeev-type integral equations
\begin{equation}
|\psi_{\alpha} \rangle= G_\alpha^{(l)} (z)
{\bf 1}^{N}_\alpha v^{(s)}_\alpha \sum_{\gamma\neq\alpha}
{\bf 1}^{N}_\gamma |\psi_{\gamma} \rangle,
\label{feqsapp}
\end{equation}
i.e.\ the short-range potential
$v_\alpha^{(s)}$ in the three-body
Hilbert space is taken to have a separable form, viz.
\begin{equation}
v_\alpha^{(s)}\approx \sum_{n,\nu ,n^{\prime },
\nu ^{\prime }=0}^N|\widetilde{n\nu }\rangle _\alpha \;
\underline{v}_{\alpha \beta }^{(s)}
\;\mbox{}_\beta \langle \widetilde{n^{\prime }
\nu ^{\prime } }|,  \label{sepfe}
\end{equation}
where $\underline{v}_{\alpha \beta}^{(s)}=
\mbox{}_\alpha \langle n\nu |
v_\alpha^{(s)}|n^{\prime }\nu ^{\prime} \rangle_\beta$.
In Ref.\ \cite{pzsy} the validity of the approximation
were proved. The argumentation in Ref.\ \cite{pzsy} relies on 
the square integrable
property of the term $v_\alpha^{(s)} |\psi_\gamma \rangle$, 
$\gamma \neq \alpha$. Thus this approximation is
justified also for complex energies as long as this property remains valid.
In Eq.~(\ref{sepfe}) the ket and bra states are defined
for different fragmentation, depending on the
environment of the potential operators in the equations.
Now, with this approximation, the solution of the homogeneous
Faddeev-Merkuriev equations
turns into solution of matrix equations for the component vector
$\underline{\psi}_{\alpha}=
 \mbox{}_\alpha \langle \widetilde{ n\nu } | \psi_\alpha  \rangle$
\begin{equation}
\underline{\psi}_{\alpha} = \underline{G}_\alpha^{(l)} (z)
\underline{v}^{(s)}_\alpha \sum_{\gamma\neq\alpha}
\underline{\psi}_{\gamma},
\label{feqm}
\end{equation}
where $\underline{G}_\alpha^{(l)}=\mbox{}_\alpha \langle \widetilde{
n\nu } |G_\alpha^{(l)}|\widetilde{n^{\prime}\nu^{\prime}
 }\rangle_\alpha$. A unique solution exists if
and only if
\begin{equation}
\det \{ [ \underline{G}^{(l)}(z)]^{-1} - \underline{v}^{(s)} \} =0.
\end{equation}

The Green's operator ${G}_\alpha^{(l)}$ is a solution of
the auxiliary three-body problem with the Hamiltonian ${H}_\alpha^{(l)}$.
To determine it uniquely
one should start again from Faddeev-type integral equations, which does not
seem to lead any further, or from the triad of
Lippmann-Schwinger equations \cite{glockle}.
The triad of Lippmann-Schwinger equations, although they do not possess
compact kernels and  thus they are not amenable
for practical calculations, also define the solution in an unique
way. They are, in fact, related to the adjoint representation
of the Faddeev operator \cite{sl}.
The Hamiltonian $H_\alpha^{(l)}$, however, has a peculiar
property that it supports bound state only in the subsystem $\alpha$, and
thus it has only one kind of asymptotic channel, the $\alpha$ channel.
For such a system one single Lippmann-Schwinger equation is sufficient for
an unique solution \cite{sandhas}.

The appropriate equation takes the form
\begin{equation}
G_\alpha^{(l)}=\widetilde{G}_\alpha +
\widetilde{G}_\alpha  U^\alpha G_\alpha^{(l)},
\label{lsgc}
\end{equation}
where $\widetilde{G}_\alpha$ is the resolvent 
channel-distorted long-range Hamiltonian,
\begin{equation}
\widetilde{H}_\alpha=H^0+v_\alpha+u_\alpha^{(l)},
\label{htilde}
\end{equation}
and  $U^\alpha=W + v_\beta^{(l)}+v_\gamma^{(l)}  -u_\alpha^{(l)}$.
The auxiliary potential
$u_\alpha^{(l)}$ depends on the coordinate $y_\alpha$. It is defined  such
that it does not support any bound state and  has the asymptotic form
$u_\alpha^{(l)} \sim {e_\alpha (e_\beta+e_\gamma) }/{y_\alpha}$
as ${y_\alpha \to \infty}$. In fact, $u_\alpha^{(l)}$
is an effective Coulomb interaction between the center of
mass of the subsystem $\alpha$ (with
charge $e_\beta+e_\gamma$) and the third particle
(with charge $e_\alpha$). Its role is to compensate the Coulomb tail of the
potentials $v_\beta^{(l)}+v_\gamma^{(l)}$ in $\Omega_\alpha$.
If $u_\alpha^{(l)}$ is a repulsive Coulomb potential the requirement that it
should not support bound states can can easily be fulfilled. For attractive
$u_\alpha^{(l)}$ the infinitely many bound states should be projected out,
which leads to a non-local potential.

It is important to realize that in this approach to get the solution only
the matrix elements  $\underline{G}_\alpha^{(l)}$ are needed, i.e.
only the representation of the Green's operator
on a compact subset of the Hilbert space are required.
So, although Eq.~(\ref{lsgc}) does not possesses a compact kernel on the whole
three-body Hilbert space its matrix form is effectively a compact equation
on the subspace spanned by finite number of CS functions \cite{pzsy}.
Thus we can perform
an approximation, similar to Eq.\ (\ref{sepfe}),
on the potential $U^\alpha$ in Eq.~(\ref{lsgc}),
with bases of the same fragmentation $\alpha$
applied  on both sides of the operator. Now the integral
equation reduces to an analogous set of linear algebraic
equation with the operators replaced by their matrix representations.
The solution is given by
\begin{equation}
[\underline{G}_\alpha^{(l)}(z)]^{-1} =
[\underline{\widetilde{G}}_\alpha (z)]^{-1} - \underline{U}^{\alpha}.
\end{equation}

The most crucial point in this procedure is the
 calculation of the matrix elements
$\underline{\widetilde{G}}_{\alpha}=
\mbox{}_\alpha \langle \widetilde{n\nu } |
\widetilde{G}_\alpha |  \widetilde{ n^{\prime }\nu^{\prime}
}\rangle_\alpha $, since the  potential
matrix elements $\underline{v}^{(s)}_{\alpha \beta}$ and
$\underline{U}^{\alpha}$ can always be calculated numerically by making use of
the transformation of Jacobi coordinates.
The Green's operator $\widetilde{G}_\alpha$
is a resolvent of the sum of two commuting Hamiltonians,
$\widetilde{H}_\alpha = h_{x_\alpha}+h_{y_\alpha}$,
where $h_{x_\alpha}=h^0_{x_\alpha}+v_\alpha$ and
$h_{y_\alpha}=h^0_{y_\alpha}+u_\alpha^{(l)}$,
which act in different two-body Hilbert spaces.
Thus, using  the convolution theorem the three-body Green's operator
$\widetilde{G}_\alpha$ equates to
a convolution integral of two-body Green's operators, i.e.
\begin{equation}
\widetilde{G}_\alpha (z)=
 \frac 1{2\pi \mathrm{i}}\oint_C
dz^\prime \,g_{x_\alpha }(z-z^\prime)\;
g_{y_\alpha}(z^\prime),
 \label{contourint}
\end{equation}
where
$g_{x_\alpha}(z)=(z-h_{x_\alpha})^{-1}$  and
$g_{y_\alpha}(z)=(z-h_{y_\alpha})^{-1}$.
The contour $C$ should be taken  counterclockwise
around the continuous spectrum of $h_{y_\alpha }$
so that $g_{x_\alpha }$ is analytic in the domain encircled
by $C$.

\begin{figure}
\psfig{file=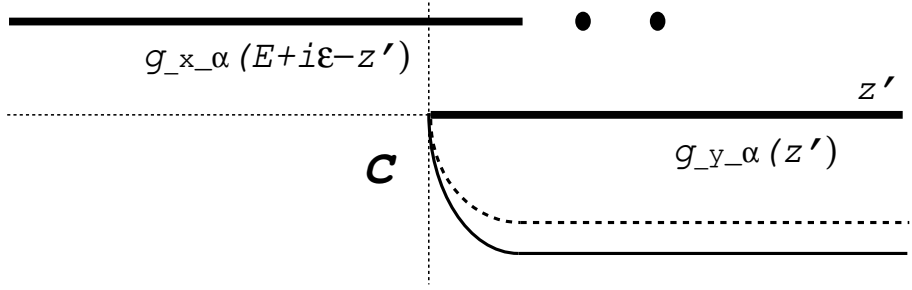,width=12cm}

\caption{Analytic structure of $g_{x_\alpha }(z-z^\prime)\;
g_{y_\alpha}(z^\prime)$ as a function of $z^\prime$ with
$z=E+{\mathrm{i}}\varepsilon$, $E>0$, $\varepsilon>0$.
The contour $C$ encircles the continuous spectrum of
$h_{y_\alpha}$. A part of it, which goes on the unphysical
Riemann-sheet of $g_{y_\alpha}$, is drawn by broken line.}

\label{fig1}
\end{figure}

To examine the structure of the integrand let us
shift the spectrum of $g_{x_\alpha }$ by
taking  $z=E +{\mathrm{i}}\varepsilon$  with
positive $\varepsilon$. By doing so,
the two spectra become well separated and
the spectrum of $g_{y_\alpha}$ can be encircled.
Next the contour $C$ is deformed analytically
in such a way that the upper part descends to the unphysical
Riemann sheet of $g_{y_\alpha}$, while
the lower part of $C$ can be detoured away from the cut
 [see  Fig.~\ref{fig1}]. The contour still
encircles the branch cut singularity of $g_{y_\alpha}$,
but in the  $\varepsilon\to 0$ limit it now
avoids the singularities of $g_{x_\alpha}$.
Moreover, by continuing to negative values of  $\varepsilon$, in order that
we can calculate resonances, the branch cut and pole singularities of
$g_{x_\alpha}$ move
onto the second Riemann sheet of $g_{y_\alpha}$ and, at the same time,
the  branch cut of $g_{y_\alpha}$ moves onto the second Riemann sheet
of $g_{x_\alpha}$. Thus, the mathematical conditions for
the contour integral representation of $\widetilde{G}_\alpha (z)$ in
Eq.~(\ref{contourint}) can be fulfilled also for complex energies
with negative imaginary part.
In this respect there is only a gradual difference between the
bound- and resonant-state calculations. Now,
the matrix elements $\underline{\widetilde{G}}_\alpha$
can be cast in the form
\begin{equation}
\widetilde{\underline{G}}_\alpha (z)=
 \frac 1{2\pi \mathrm{i}}\oint_C
dz^\prime \,\underline{g}_{x_\alpha }(z-z^\prime)\;
\underline{g}_{y_\alpha}(z^\prime),
\label{contourint2}
\end{equation}
where the corresponding CS matrix elements of the two-body Green's operators in
the integrand are known analytically for all complex energies \cite{cpc},
and thus the convolution integral can be performed also in the practice.

\section{Numerical illustration}
To illustrate the feasibility of this method
we examine the convergence of the results for three-body
resonant-state energies. For this purpose we take an
Ali--Bodmer-type toy-model for the
charged three-$\alpha$ system interacting via $s$-wave
short-range interaction.
To improve its properties we add a phenomenological
three-body potential.
Adopting Noble's splitting we have
%%%
\begin{equation}
v_\alpha^{(s)}(r)=  V_{1} \exp\{ -r^2/{\beta_1}^2\} +
V_{2} \exp\{ -r^2/{\beta_2}^2\}
\label{filpot}
\end{equation}
with $V_{1}=125$ MeV, $V_{2}=-30.18$ MeV, ${\beta}_1=1.53$ fm,
${\beta}_2=2.85$ fm, and
%%%
\begin{equation}
v_\alpha^{(l)}(r)=  4 e^2/r.
\end{equation}
%%%
We use units such that $\hbar^2/m=41.47$ MeV, $e^2=1.44$ MeV fm. The mass 
of the $\alpha$-particle is chosen as $M=3.973 m$, where
$m$ denotes the mass of the nucleon.
%%%%%%%%%%%%%%%%%%%%%%%%%%
The three body potential is taken to have Gaussian form
\begin{equation}
W(\rho) =V \exp\{-\rho^2/\beta^2 \} ,
%\label{v_3}
\end{equation}
where $\rho^2 =\sum\limits_{i=1}^{3} {\bf r} _i^2 $,
$V=-31.935$ MeV and $\beta=3.315$ fm. Here
 ${\bf r}_i$ stands for the
position vector of $i$-th particle in the center of
mass frame of three-$\alpha$ system.
%%%%%%%%%%%%%%%%%%%%%%%%%%%%
We select states with total angular momentum
$L=0$. In Table I we show the convergence of the energy
of the ground-state
and of the first resonant-state with respect to $N$,
the number of CS functions employed in the expansion.
The selected resonance is the experimentally well-known
sharp state which has a great relevance in nuclear synthesis.

For comparison we have recalculated the results of Ref.\
\cite{fedorov}, where a two-channel model has been proposed for
the three-$\alpha$ system.  Table II shows a good agreement between
the two independent methods as far as the position of the
resonance is considered. For the width of the resonant state 
we got an order of magnitude less which 
indicate an order of magnitude longer lifetime.
The origin of the discrepancy, in our opinion, could be due to the not proper
implementation of the asymptotic boundary conditions in
Ref.\ \cite{fedorov}.

\section{Conclusions}
In this article we have proposed a new method for calculating resonances
in three-body Coulombic systems. 
The homogeneous Faddeev-Merkuriev integral equations were solved for 
complex energies. For this, being an integral equation approach, the
no boundary conditions are needed.
The procedure is an extension of the
well-established Coulomb-Sturmian separable expansion approach
\cite{pzwp,pzsc,pzatom}.
In fact, the extension is nothing else but the proper analytic continuation
of the contour integral representation of the Green's operator.
In the non-Coulomb case the method is readily applicable, only
the Coulomb Green's operators have to be replaced by the free one.

\section*{Acknowledgments}
This work has been supported by OTKA under Contracts No.\ T026233
and No.\ T029003, by Russian Foundation for Basic Research
Grant No. 98-02-18190 and partially by the exchange program
between the Hungarian and the Russian Academies of Sciences.

\begin{table}
\caption{Convergence of the ground-state and of the first resonant-state
 energy (in MeV) of a three-$\alpha$ system
interacting via the potential of (\ref{filpot})
with increasing basis for the separable expansion.
$N$ denotes the maximum number of
basis states employed for $n$ and $\nu$ in Eq. (\ref{sepfe}).
}
\label{tabc}
\begin{tabular}{rll}
\hline
$N$ & \multicolumn{1}{c}{\mbox{$E$} } &
\multicolumn{1}{c}{\mbox{$E=E_r-\mbox{i}\Gamma /2$} }  \\
\hline
15 & -7.283686\phantom{000} & 0.3859108 -i 0.000011 \\
16 & -7.283744 & 0.3854244 -i 0.000011 \\
17 & -7.283779 & 0.3851242 -i 0.000011 \\
18 & -7.283801 & 0.3849323 -i 0.000012 \\
19 & -7.283815 & 0.3848056 -i 0.000012 \\
20 & -7.283824 & 0.3847236 -i 0.000012 \\
21 & -7.283829 & 0.3846683 -i 0.000012 \\
22 & -7.283833 & 0.3846308 -i 0.000012 \\
23 & -7.283836 & 0.3846053 -i 0.000012 \\
24 & -7.283837 & 0.3845873 -i 0.000013 \\
25 & -7.283838 & 0.3845748 -i 0.000013 \\
26 & -7.283839 & 0.3845658 -i 0.000013 \\
27 & -7.283840 & 0.3845593 -i 0.000013 \\
28 & -7.283840 & 0.3845546 -i 0.000013 \\
29 & -7.283640 & 0.3845512 -i 0.000013 \\
\hline
\end{tabular}
\end{table}

\begin{table}
\caption{Results for the three-$\alpha$ model of Ref.\
\cite{fedorov} (in MeV).}
\label{tabd}
\begin{tabular}{llll}
  & \multicolumn{1}{c}{\mbox{$E$} }
 & \multicolumn{2}{c}{\mbox{$E=E_r-\mbox{i}\Gamma /2$} }  \\
\hline
This work             & -6.8053\phantom{0000}    & 0.3572  & - i 0.000002  \\
Ref.\ \cite{fedorov}  & -6.81      & 0.38    &- i 0.000020 \\
\hline
\end{tabular}
\end{table}

{}
%\end{references}


\begin{thebibliography}{99}
%\begin{references}

\bibitem{fm-book}  Faddeev~L.~D. and Merkuriev S.~P.:
 {\it Quantum Scattering
Theory for Several Particle Systems}, (Kluver, Dordrech), (1993).

\bibitem{pzwp}  Papp~Z. and Plessas~W.: Phys.~Rev.~C
{\bf 54}, 50 (1996).

\bibitem{pzsc}  Papp~Z.:  Phys.~Rev.~C {\bf 55}, 1080 (1997).

\bibitem{pzatom}  Papp~Z.: Few-Body~Systems, {\bf 24}, 263 (1998).

\bibitem{pzsy}  Papp~Z. and Yakovlev~S.~L.: submitted, (nucl-th/9903078).

\bibitem{galindo} Alvarez-Estrada~R.~F.\ and Galiondo ~A.:
Nuovo Cim. {\bf B44}, 47 (1978).

\bibitem{noble}  Noble~J.~V.: Phys.~Rev.\ {\bf 161}, 945 (1967).

\bibitem{vanzani} Vanzani~V.:
{\it Few-Body Nuclear Physics}, (IAEA Vienna), 57 (1978).

\bibitem{glockle}  Gl\"ockle~W.: Nucl.~Phys.\ A {\bf 141}, 620 (1970).

\bibitem{sl}  Yakovlev~S.~L.: Theor.~Math.~Phys.\ {\bf 102}, 323 (1995);
{\bf 107}, 513 (1996).

\bibitem{sandhas} Sandhas~W.:
  {\it Few-Body Nuclear Physics}, (IAEA Vienna), 3 (1978).

\bibitem{cpc} Papp~Z.: J.~Phys.\  A {\bf 20}, 153 (1987);
 Phys.~Rev.~C {\bf 38}, 2457 (1988);
 Phys.~Rev.~A {\bf 46}, 4437 (1992);
 Comp.~Phys.~Comm.\  {\bf 70}, 426 (1992); ibid {\bf 70}, 435 (1992);
 K\'onya~B., L\'evai~G. and Papp~Z.:  J.\ Math.\ Phys.
{\bf 38}, 4832 (1997).

\bibitem{fedorov}  Fedorov~D.~V. and  Jensen~A.~S.:\ Phys\. Lett.\ 
{\bf B389} 631 (1996)
\end{thebibliography}
\end{document}